\documentclass[prl,twocolumn,superscriptaddress,showpacs,floatfix]{revtex4}
\input epsf
\usepackage{graphicx,amsmath,amssymb}
\begin{document}
\title
{No-soliton--soliton phase transition in a trapped one-dimensional Bose gas}
\date{\today}

\author{Vanja Dunjko}
\affiliation{
Department of Physics \& Astronomy,
University of Southern California,
Los Angeles, CA 90089-0484, USA
}
\author{Christopher P. Herzog}
\affiliation{Department of Physics, Princeton University, Princeton, NJ 08544,
 USA}
\affiliation{Ecole Normale Sup\'{e}rieure, Laboratoire Kastler-Brossel, 24 Rue
Lhomond, 75231 Paris Cedex 05, France}
\author{Yvan Castin}
\affiliation{Ecole Normale Sup\'{e}rieure, Laboratoire Kastler-Brossel, 24 Rue
Lhomond, 75231 Paris Cedex 05, France}
\author{Maxim Olshanii}
\email[email: ]{olshanii@physics.usc.edu}
\affiliation{
Department of Physics \& Astronomy,
University of Southern California,
Los Angeles, CA 90089-0484, USA
}
\affiliation{Ecole Normale Sup\'{e}rieure, Laboratoire Kastler-Brossel, 24 Rue
Lhomond, 75231 Paris Cedex 05, France}
\affiliation{Institute for Theoretical Atomic and Molecular Physics,
Harvard-Smithsonian Center for Astrophysics, Cambridge, MA 02138, USA}
\begin{abstract}
Following the experimental observation of bright matter-wave
solitons [L. Khaykovich \textit{et al.}, Science \textbf{296},
1290 (2002); K. E. Strecker \textit{et al.}, Nature (London)
\textbf{417,} 150 (2002)], we develop a semi-phenomenological
theory for soliton thermodynamics and find the condensation
temperature. Under a modified thermodynamic limit, the condensate
occupation at the critical temperature undergoes a sudden jump to
a nonzero value, indicating a discontinuous phase transition.
Treating the condensation as a diffusion over a barrier shows that
the condensation time is exponentially long as one approaches the
thermodynamic limit, and the longest near the critical
temperature.
\end{abstract}
\maketitle
%
%
Quantum-degenerate atomic gases are attracting attention as
versatile physics laboratories with unprecedented level of
experimental control \cite{Ketterle}. Nonlinear science in
particular should benefit from the study of dark \cite{dark} and
bright \cite{Salomon, Hulet} matter-wave solitons
\cite{early_soliton} recently observed in one-dimensional (1D)
Bose gases, whose mean-field dynamics is described by the 1D
nonlinear Schr\"{o}dinger equation. Much recent theoretical work
on 1D matter-wave bright solitons is dedicated to explaining their
formation, observed number, and dynamics \cite{soliton_dynamics}
and to proposing new ways of generating them
\cite{new_soliton_generation}. We shall consider instead the
thermodynamics of this system, rarely of central importance in
other nonlinear media.

Microscopically, a bright soliton is a many-body well-localized
bound state of an attractive gas. While at zero temperature this
state contains all the atoms, at sufficiently high temperature it
should disappear. We describe the 1D harmonically trapped gas
using a model with exactly solvable thermodynamics and show that,
in a modified thermodynamic limit, the change in the bound-state
population happens through a discontinuous phase transition. Our
main computational result is the critical temperature, with
accompanying phase diagram. Finally, we turn to metastability
issues and consider condensate formation (and evaporation, above
the critical temperature) as a diffusion over a potential barrier.
We find that the transition time increases roughly exponentially
with a parameter characterizing the proximity to the thermodynamic
limit, the duration peaking near the critical temperature.
%
\paragraph*{Ground state.}
The Hamiltonian for a 1D system of $N$ attractive $\delta$-interacting
bosons in a harmonic trap is
%
\begin{equation}
\widehat{H} = -\frac{\hbar^2}{2m} \sum_{i=1}^{N}
\frac{\partial^2}{\partial z_{i}^2}
+\, g_{\text{1D}}
\!\!\!\!\!\!\!
\sum_{1 \le i < j \le N}
\!\!\!\!\!\!
\delta(z_{j}-z_{i})
+ \sum_{i=1}^{N}
\!\!\!\!\!\!
\underset{
         \stackrel{\ }{\rm \quad\; trap}
         }{U}
\!\!\!\!\!\!(z_{i})
\, .
\label{eq:Hamiltonian}
\end{equation}
%
We assume the 1D coupling constant to be negative, $g_{\text{1D}}
< 0$; its relation to the 1D scattering length $a_{\text{1D}}$ is
$g_{\text{1D}}=-2\hbar^{2}/(m\,a_{\text{1D}})$. In the actual
waveguide or cigar trap experiments the value of $a_{\text{1D}}$
is governed by the 3D scattering length $a$ and size of the ground
transverse state $a_{\perp}=\sqrt{2\hbar / m \,\omega_{\perp}}$,
namely $ a_{\text{1D}} = -(a^{2}_{\perp}/2 a)\,[1-{\mathcal
C}(a/a_{\perp})] $, where $ {\mathcal C} =\,
|\zeta(1/2)| = 1.4603 \dots
\, $\cite{Maxim_1D}. Here $\omega_{\perp}$ is the frequency of the
transverse confinement, $m$ is the atomic mass, and $\zeta(x)$ is
the Riemann zeta-function.

In the absence of the trapping potential the exact ground state of
the Hamiltonian (\ref{eq:Hamiltonian}) assumes a known Bijl-Jastrow form
\cite{first_soliton,Chris,Kolomeisky}
%
$
\Psi(z_{1},\,\ldots,\,z_{N}) =
A \prod_{1 \le i < j \le N} e^{-|z_{j}-z_{i}|/a_{\text{1D}}}
\, ,
%
$
%
where $A$ is the normalization constant.
Ground state energy is then
%
$
E_{0}^{\text{exact}}(N) =
-{\scriptstyle{\frac{1}{24}}} \, \left(m g_{\text{1D}}^2/\hbar^2\right)
\, N(N^2-1)
\, .
$
%
The center of mass of a system in this state is delocalized,
although the relative motion of particles is tightly bound. If we
now add a weak harmonic confinement $U_{\text{trap}}(z) =
m\omega_{z}^2 z^2/2 \,\, ,$ (where the chemical potential of the
free space soliton $m(gN)^2/\hbar^2 \gg \hbar\omega_{z}$), the
one-body density distribution becomes localized around the origin.
In the limit of large $N$, mean-field theory predicts a one-body
density distribution of the form
%
$
\rho(z) = \rho_{0}/\cosh^{2}(z/\ell) \, ,
$
%
where
$
\rho_{0} = (2\ell)^{-1}
$
is the peak density and
$
\ell = 2\hbar^2/mgN
$
is the density profile width \cite{Chris,Kolomeisky}.

%
\paragraph*{Model Hamiltonian.}
We analyze the thermodynamics of our system using the following Hamiltonian:
\begin{eqnarray}
\widehat{\cal H} =
E_{0}(\widehat{N}_{0})
+ \sum_{s=0}^{\infty} \epsilon_{s}\,
\hat{a}_{s}\text{}^{\!\!\!{\scriptscriptstyle\dagger}} \hat{a}_{s}
\, ,
\label{eq:soliton_Hamiltonian}
\end{eqnarray}
where
%
$
E_{0}(N_{0})
=
-{\scriptstyle{\frac{1}{24}}} \, \left(m g_{\text{1D}}^2/\hbar^2\right)
\,N_{0}^3
$
%
is the soliton energy for $N_{0} \gg 1$,
$
\widehat{N}_{0}
=
N - \widehat{N}^{\prime}
$
is the number of particles in the
soliton, $N$ is the fixed total number of particles,
%
$
\widehat{N}^{\prime} =
\sum_{s=0}^{\infty} \,\hat{a}_{s}\text{}^{\!\!\!{\scriptscriptstyle\dagger}}
\hat{a}_{s}
$
%
is the number of noncondensed particles, $\epsilon_{s} =
\hbar\omega_{z}s$ is the one-body energy spectrum of the 1D
harmonic oscillator, and $\omega_{z}$ is the frequency
characterizing the longitudinal confinement. The Fock vacuum for
the bosonic annihilation and creation operators $\hat{a}_{s}$ and
$\hat{a}_{s}\text{}^{\!\!\!{\scriptscriptstyle\dagger}},$ where
$[\hat{a}_{s},\,
\hat{a}_{s'}\text{}^{\!\!\!\!\!{\scriptscriptstyle\dagger}\,}] =
\delta_{s,s'},$ is the harmonic oscillator ground state. Note that
the ground state of the Hamiltonian $\widehat{\cal H}$ corresponds
to all $N$ particles being in the soliton, so that the first
excited state is $N-1$ particles in the soliton, $1$ particle in
the lowest harmonic oscillator state $s=0.$ The energy spectrum
thus has a gap, about equal to the chemical potential of the free
space soliton, as can also be confirmed by Bogoliubov analysis
\cite{Chris}.

\begin{figure}
\epsfxsize=0.42\textwidth
\begin{center}\epsfbox{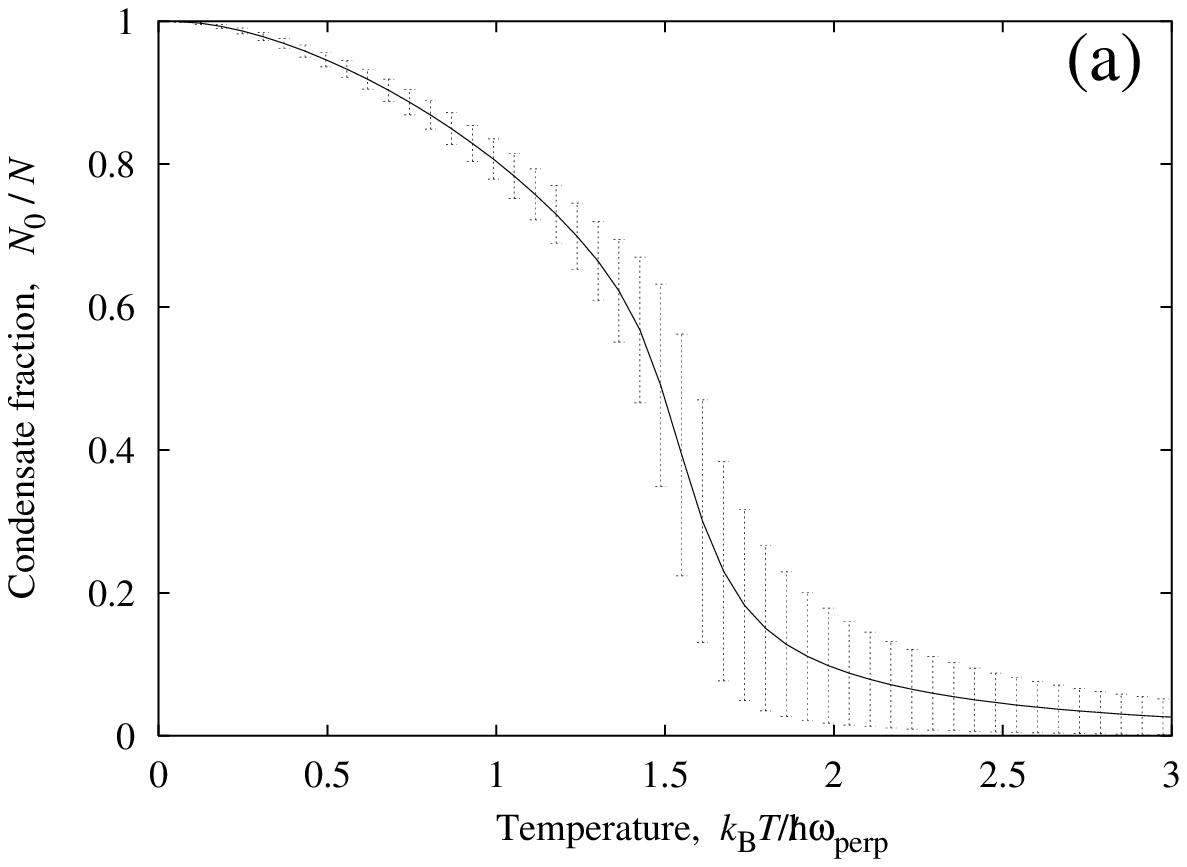}\end{center}
\epsfxsize=0.42\textwidth
\begin{center}\epsfbox{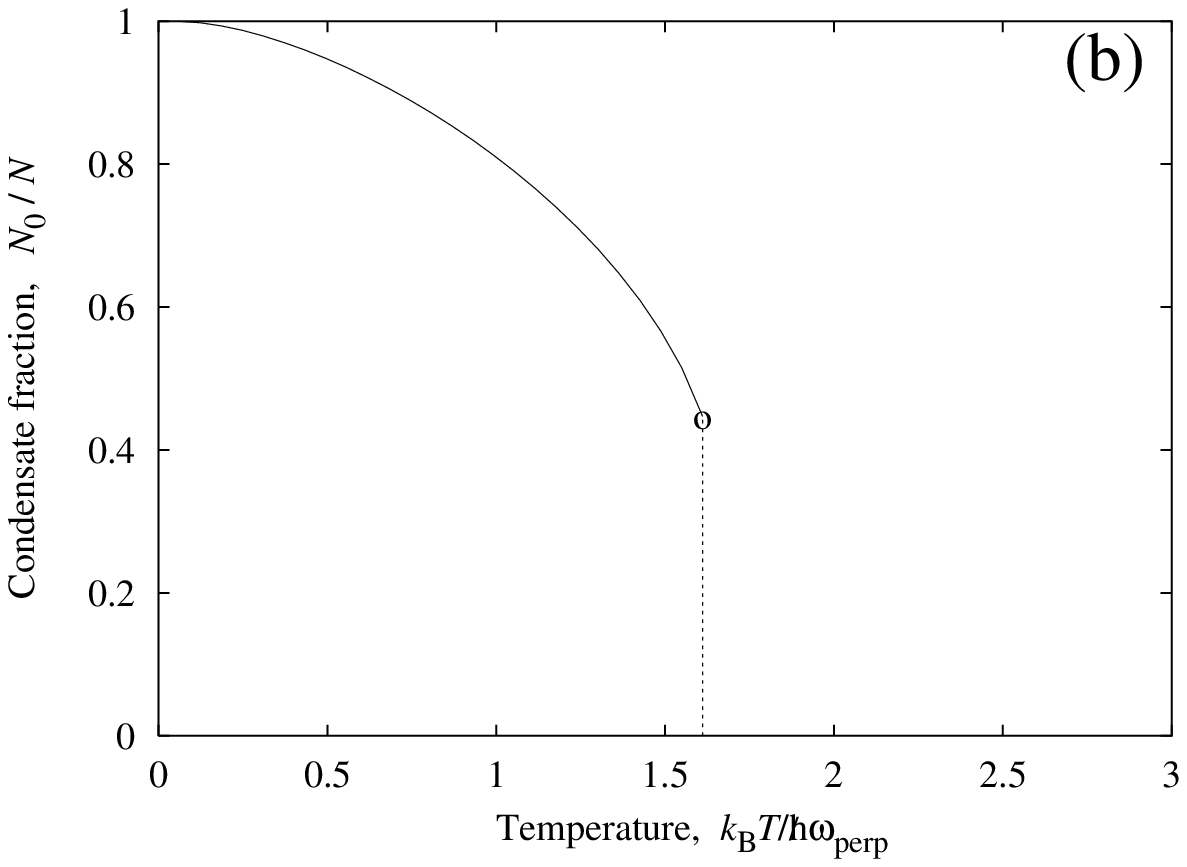}\end{center}
\caption { Solitonic condensate fraction vs.\ temperature for $N
=$ 1050 $^7$Li atoms of a scattering length $a_{3D} =$ -1.46 nm,
confined in a cigar-shape trap with frequencies $\omega_{\perp} =$
2$\pi$ $\times$ 300 Hz and $\omega_{z} =$ 2$\pi$ $\times$ 3 Hz.
This set of parameters corresponds to $\sigma = 56 \,\,
\hbar\omega_{\perp}$, $\varepsilon = 1.77 \times 10^{-4}$, and
$\eta = 0.186$. (a) Finite theory; (b) thermodynamic limit, $
\varepsilon \to 0$, with the value of the energy scale $\sigma$
and rescaled total number of particles $\eta$ kept constant. The
critical temperature $k_{\text{B}}T_{\text{c}} = 1.65\,\,
\hbar\omega_{\perp}$ corresponds to $\tau_{\text{c}} = .0293$. }
\label{fig:N0_vs_T}
\end{figure}

By far the most ``phenomenological'' assumption we made is the
conjecture that even at finite temperatures the condensate is not
fragmented, but only depleted. This conjecture is confirmed in
\textit{ab initio} calculations for mesoscopic ($N \sim 40$)
numbers of atoms in a 1D box \cite{Chris_report}. Next comes the
assumption that the solitonic condensate is not deformed by the
external harmonic confinement, leading to the
requirement\footnote{ One may improve this approximation by using
the soliton energy obtained via numerical solving a
Gross-Pitaevskii equation in presence of the harmonic trap.}
%
$
\hbar\omega_{z} \ll \partial E_{0}/\partial N_{0}
\sim m(gN_{0})^2/\hbar^2
\, .
%
$
%
We further neglect interactions both between the noncondensed
particles and between noncondensed particles and the soliton. The
former can be justified for the case of well-delocalized
noncondensed cloud. To justify the latter notice that under the
above requirement of no deformation by the external confinement,
the size of the soliton $\ell$ becomes much less than the size of
the trap ground state $a_{z}=\sqrt{2\hbar / m \,\omega_{z}}$. One
can show that strong but well-localized perturbation cannot affect
the overall density of states. The assumption can also be
justified via a Bogoliubov-like analysis of the problem
\cite{Kolomeisky,Chris_report}. Finally, in our Hamiltonian we
neglect the center-of-mass motion of the soliton as it can be
shown to be fully decoupled from the rest of the system.
\paragraph*{Finite size thermodynamics.}
Thermodynamics of our system is determined by the canonical
partition function
%
%
$\left.\right.\qquad
{\cal Q}(N,\, T) = \sum_{N_{0} = 0}^{N}
e^{-\frac{E_{0}(N_{0})}{k_{\text{B}}T}}
Q^{\prime}(N-N_{0}, T)
,
%
$
%
\\ where \\
$
Q^{\prime}(N^{\prime},\, T) =
\underset{\scriptstyle{\{n_{s}\},\,\sum_{s}
n_{s}=N^\prime}}{\!\!\!\!\!\!\!\!\!\!\!\!\!\!\sum}
\!\!\!\!\!\!\!\!\!\!\!\!\!\!\!\!\!\! e^{-\sum_{\scriptstyle{s}}
\frac{\scriptstyle{\epsilon_{s}
n_{s}}}{\scriptstyle{k_{\text{B}}T}}}
=\prod_{q=1}^{N^{\prime}}
\Bigl(\displaystyle{1-
e^{\textstyle{-\frac{\hbar\omega_{z}}{k_{\text{B}}T}q}}}\Bigr)^{-1}
$
is the canonical partition function for $N^{\prime}$
non-interacting bosons in a 1D harmonic trap \cite{Chris_canon}.
In Fig. \ref{fig:N0_vs_T}(a) we show the mean soliton occupation
and its standard deviation due to thermodynamic fluctuations as a
function of temperature for a typical experimental setup. In this
calculation we have replaced the sum over $N_{0}$ by an integral
and used a large temperature $(k_{\text{B}}T \gg \hbar\omega_{z})$
approximation for the partition function of noncondensed
particles,
$
%
Q^{\prime}(N^{\prime},\, T)
\!\!
\stackrel{
          \frac{\hbar\omega_{z}}{k_{\text{B}}T} \ll 1
         }{\stackrel{\ }{\approx}}
\!\!
\exp\left(
\frac{\scriptstyle{k_{\text{B}}T}}{\scriptstyle{\hbar\omega_{z}}}
\left[ {\text{Li}}_{2}(1)-{\text{Li}}_{2}(
e^{-\frac{N^{\prime}\hbar\omega_{z}}{k_{\text{B}}T}}) \right]
\right)
,
$
where ${\text{Li}}_{2}(\zeta)=\int_{\zeta}^{0}\frac{\ln (1-t)}{t}\,dt$
is the dilogarithm function.
%
%
%
\paragraph*{Scaling.} The following scaling has proven useful for
analyzing thermodynamical properties of our system. Let us
introduce a dimensionless parameter $\varepsilon \equiv (m
g_{\text{1D}}^2)/(\hbar^3 \omega_{z})$, a dimensionless rescaled
temperature $\tau \equiv k_{\text{B}}T/\sigma,$ and a
dimensionless rescaled number of particles $\eta \equiv
\varepsilon N$, where $\sigma \equiv (\hbar^4\omega_{z}^2)/(m
g_{\text{1D}}^2)$. We also introduce rescaled numbers of condensed
($\eta_{0}$) and noncondensed ($\eta^{\prime}$) particles in an
analogous way. The parameter $\varepsilon$ can be shown to be
proportional to the ratio between the binding energy of a 1D dimer
$\epsilon_{b} = (m g_{\text{1D}}^2)/(4\hbar^2)$ and the trap level
spacing $\hbar\omega_{z}$: $\varepsilon = (1/4) \,
(\epsilon_{b}/\hbar\omega_{z})$. The rescaled partition function
is
\begin{eqnarray}
{\cal Q}(\eta,\, \tau)
&=&
\varepsilon
\int_{0}^{\eta} d\!\eta_{0} \,
e^{-f_{\varepsilon}(\eta,\, \tau \, | \, \eta_{0})/\tau} \,; \notag \\
%
f_{\varepsilon}(\eta,\, \tau \, | \, \eta_{0})
&=&
- {\textstyle \frac{1}{\varepsilon}}
{\textstyle
\left\{
{\scriptstyle{\frac{1}{24}}} \, \eta_{0}^3
+\!\tau^2 \left[ {\text{Li}}_{2}(1)-{\text{Li}}_{2}(
e^{-\frac{\eta-\eta_{0}}{\tau}}) \right]
\right\}
}
 \, . \notag
\end{eqnarray}
Notice that the positions of the extrema of the Landau free energy
$f_{\varepsilon}(\eta,\, \tau \, | \, \eta_{0})$ as a function of
$\eta_{0}$ does not depend on either the energy scale $\sigma$ or
the parameter $\varepsilon$. Here we attempt to identify the
rescaled number of condensed particles $\eta_{0}$ with the order
parameter of a possible phase transition. The distribution of the
order parameter is given by the Boltzmann distribution of a
particle with coordinate $\eta_{0}$ moving in the potential field
$f_{\varepsilon}(\eta,\, \tau \, | \, \eta_{0})$ at a temperature
$\tau$. In the limit where the variation of the effective
potential $f_{\varepsilon}$ becomes large as compared to the
temperature, the order parameter becomes frozen at the global
minimum of $f_{\varepsilon}$. The appearance of a global minimum
different from $\eta_{0}=0$ is associated with a phase transition.
Depending on whether the minimum undergoes a finite jump from zero
or gradually moves away from it as a function of temperature, the
phase transition is usually said to be first or second order.

\begin{figure}
\epsfxsize=0.42\textwidth
\begin{center}\epsfbox{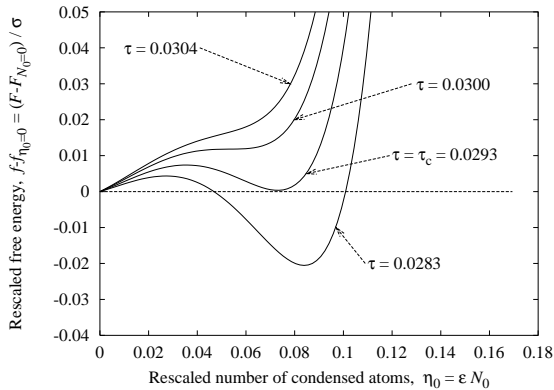}\end{center}
\caption { Free energy as a function of the order parameter
$\eta_{0}$ for the set of parameters corresponding to Fig. 1. }
\label{fig:free_energy}
\end{figure}

Note that the shape of the free energy $f_{\varepsilon}$ as a
function of $\eta_{0}$ is given by a universal function
independent of both the energy scale $\sigma$ and the parameter
$\varepsilon$. However, the amplitude of $f_{\varepsilon}$ is
inversely proportional to $\varepsilon$.
%
%
\paragraph*{Free energy.}
Let us analyze the evolution of the free energy vs.\ $\eta_{0}$
curve with changing temperature. Figure \ref{fig:free_energy}
corresponds to the experimental conditions of Fig.
\ref{fig:N0_vs_T}. For this set of parameters we get $\varepsilon
= 1.77 \times 10^{-4}$ and $\eta = 0.186$. For high temperatures
the only minimum of free energy is located at $\eta_{0}=0$. At
$\tau = .0300$ a second minimum appears and at $\tau =
\tau_{\text{c}} = .0293$ it becomes global ($\eta_{0}=0.0820$,
$N_{0} = 462$), moving then gradually toward $\eta_{0}=\eta$ as
the temperature decreases. Since the height of the free energy
curve is finite, we do not expect any sharp change at $\tau =
\tau_{\text{c}}$ ($k_{\text{B}}T = 1.65\, \hbar\omega_{\perp}$).
Nevertheless, Fig. \ref{fig:N0_vs_T}(a) shows that at this
temperature the condensate occupation does begin to increase
rapidly.
%
%
\paragraph*{Thermodynamic limit and phase transition.} We are going to identify
a limit where
the above-described quasi-discontinuity in the number of condensed
particles becomes a real discontinuity. Consider the following
limit:
\begin{eqnarray}
\varepsilon \to 0, \quad \tau = \text{const}, \quad \eta =
\text{const}\, . \label{thermodynamic_limit}
\end{eqnarray}
Notice that in this limit the Landau free energy
$f_{\varepsilon}(\eta,\, \tau \, | \, \eta_{0})$ grows to infinity
while preserving its shape along with positions of minima. A
typical condensate occupation vs.\ temperature dependence in the
thermodynamic limit is shown in Fig. \ref{fig:N0_vs_T}(b)
\footnote{We chose parameters for this plot so that $T_{\text{c}}$
is slightly higher than the transverse level spacing. At least at
the Boltzmann equation level, 1D gas cannot thermalize by itself.
An additional reservoir is required, thermal transverse
excitations being a good candidate. In this case the numbers of
atoms involved should be understood as those in the ground
transverse mode of the trap.}.
%
\begin{figure}
\epsfxsize=0.42\textwidth
\begin{center}\epsfbox{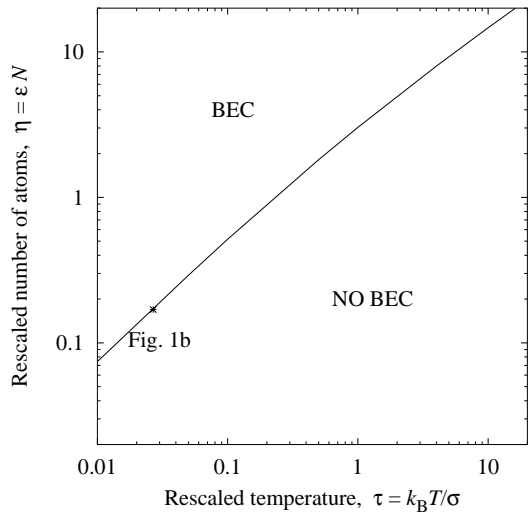}\end{center}
\caption{Phase diagram of the solitonic Bose condensate}
\label{fig:phase_diagram}
\end{figure}
%
Figure \ref{fig:phase_diagram} shows our system's phase diagram.
It is universal, using the rescaled quantities $\tau$ and $\eta$.
Our \textit{main result} is the following implicit equation for
the critical temperature $T_{\text{c}}$:
\begin{eqnarray}
N
\stackrel{\stackrel{\stackrel{\scriptstyle \tau \ll 1}{\ }}{\stackrel{
\scriptstyle \eta \ll 1}{\ }}}{=}
\frac{k_{\text{B}}T_{\text{c}}}{\hbar\omega_{z}}
\ln\left(\frac{24}{(3-w)w^2}\frac{\hbar^4 \omega_{z}^2}{m
g_{\text{1D}}^2 k_{\text{B}}T_{\text{c}}}\right) \, . \label{Tc}
\end{eqnarray}
Here $w=3+W_{0}\left(-3e^{-3}\right) = 2.82144 \ldots$, where
$W_{0}(\cdot)$ is the principal branch of the product log (Lambert
$W$) function \cite{Corless96a}, the inverse of $x \mapsto xe^{x}$
for $x \geqslant -1$. Eq. (\ref{Tc}) is valid in the currently
experimentally accessible range $\tau,\eta \ll 1$. There
$\tau_{\text{c}}$ satisfies
%
$
\eta = \tau_{\text{c}}
\ln(\frac{24}{(3-w)w^2}\frac{1}{\tau_{\text{c}}}) +
\mathcal{O}(\tau_{\text{c}}^{2})\, .
$
%
For the experiment \cite{Salomon},
%
%
we have $T_{\text{c}} \sim 1 \mu\text{K}\,\left.\right.$
\footnote{This temperature is in fact many times higher than the
transverse confinement level spacing
$\hbar\omega_{\perp}/k_{\text{B}}$, making our 1D model invalid at
$T_{\text{c}}$. In reality the experimental temperature was much
less than $T_{\text{c}}$.}.
\newcommand{\w}[3]{w{\scriptstyle (#2 #3 #1)}}
\newcommand{\J}[2]{J{\scriptstyle \uparrow}\begin{subarray}{c}{
\scriptscriptstyle #1}
\\ {\scriptscriptstyle #2} \end{subarray}}
\newcommand{\peq}[1]{p_{\scriptscriptstyle #1}^{\text{eq}}}
\newcommand{\p}[1]{p_{\scriptscriptstyle #1}}
\newcommand{\pdot}[1]{\dot{p}_{\scriptscriptstyle #1}}
\paragraph*{Metastability.} Let $\p{N_{0}}(t)$ be the probability
distribution over $N_{0}$, $\Pi_{\text{L}}(t) =
\sum_{i=0}^{N_{0}^{\text{max}}-1} \p{i}(t)$ and $\Pi_{\text{R}}(t)
= \sum_{i=N_{0}^{\text{max}}}^{N} \p{i}(t)$, where
$\varepsilon\,N_{0}^{\text{max, min}}=\eta_{0}^{\text{max, min}}$
are the local maximum and nonzero minimum of $f_{\varepsilon}$.
For temperatures around $T_{\text{c}}$ and small $\varepsilon$,
the equilibrium probability distribution $\peq{N_{0}}$ consists of
a peak at $N_{0}=0$ ($\Pi_{\text{L}}^{\text{eq}}$) and a peak at
$N_{0}^{\text{min}}$ ($\Pi_{\text{R}}^{\text{eq}}$). The latter
dominates below $T_{\text{c}}$; the former, above it. In terms of
the analogy to a particle in a potential, switching $N_{0}$ from
one peak to the other involves the system having to overcome the
potential barrier at $N_{0}^{\text{max}}$. We assume that the two
peaks satisfy the rate equation
$
\dot{\Pi}_{\text{R}}=-\dot{\Pi}_{\text{L}}=\tfrac{1}{\chi}
\left[\Pi_{\text{R}}^{\text{eq}}\, \Pi_{\text{L}} -
\Pi_{\text{L}}^{\text{eq}}\, \Pi_{\text{R}}\right]\, ,
$
whose solution is
$
\Pi_{\text{R,\,L}}(t)=
\Pi_{\text{R,\,L}}^{\text{eq}}\big(1-e^{-t/\chi}\,\big)
+\Pi_{\text{R,\,L}}(0)\,e^{-t/\chi}.
$
Below $T_{\text{c}}$, $\chi/\Pi_{\text{R}}^{\text{eq}} \approx
\chi$ is the condensation time; above it,
$\chi/\Pi_{\text{L}}^{\text{eq}} \approx \chi$ is the evaporation
time. To obtain $\chi$, we take that $\p{N_{0}}$ satisfies a rate
equation of the form
$
\pdot{N_{0}}= \w{N_{0}+1}{N_{0}}{\gets}\p{N_{0}+1}
+\w{N_{0}}{N_{0}-1}{\to}\p{N_{0}-1}
-\big[\w{N_{0}}{N_{0}-1}{\gets}+\w{N_{0}+1}{N_{0}}{\to}\big]\p{N_{0}}
$ and that the detailed balance property
$\w{N_{0}}{N_{0}-1}{\to}\peq{N_{0}-1}=\w{N_{0}}{N_{0}-1}{\gets}\peq{N_{0}}$
holds. For sharp peaks we can relate the microscopic and
macroscopic quantities through
$\Pi_{\text{L}}/\Pi_{\text{L}}^{\text{eq}} \approx \p{0}/\peq{0}$
and $ \Pi_{\text{R}}/\Pi_{\text{R}}^{\text{eq}} \approx
\p{N_{0}^{\text{min}}}/\peq{N_{0}^{\text{min}}}$. This eventually
yields \cite{van_Kampen}
\begin{equation}
\chi=\frac{1}{\Gamma \varepsilon^{2}}
\frac{S_{\text{L}}\,S_{\text{R}}}{S_{\text{L}}+S_{\text{R}}}\,
S_{\text{C}}\, , \label{chi}
\end{equation}
where\\
$S_{\text{L}}=\int_{0}^{\eta_{0}^{\text{max}}}
e^{-\frac{1}{\tau}f_{\varepsilon}(\eta,\,\tau\,|\,\eta_{0})}\,d\eta_{0}
\sim \frac{\varepsilon}{\phi'_{\eta,
\tau}(0)}\,e^{-\frac{1}{\varepsilon}\phi_{\eta, \tau}(0)}\, ,$\\
$S_{\text{R}}=\int_{\eta_{0}^{\text{max}}}^{\eta}
\! e^{-\frac{1}{\tau}f_{\varepsilon}(\eta,\, \tau \, | \,
\eta_{0})}\,d\eta_{0}
 \sim
\sqrt{
 \frac{2\pi\varepsilon}{\phi''_{\eta,
\tau}(\eta_{0}^{\text{min}})} }
%
\,e^{-\frac{1}{\varepsilon}\phi_{\eta,
\tau}(\eta_{0}^{\text{min}})}$,\\
$S_{\text{C}}=\int_{0}^{\eta_{0}^{\text{min}}}
e^{\frac{1}{\tau}f_{\varepsilon}(\eta,\, \tau \, | \,
\eta_{0})}\,d\eta_{0} \sim \sqrt{
 \frac{2\pi\varepsilon}{-\phi''_{\eta,
\tau}(\eta_{0}^{\text{max}})}}
\,e^{\frac{1}{\varepsilon}\phi_{\eta,
\tau}(\eta_{0}^{\text{max}})}\, ,$
and $\phi_{\eta, \tau}(\eta_{0}) \equiv \frac{\varepsilon}{\tau}
f_{\varepsilon}(\eta,\, \tau \, | \, \eta_{0})$.
Experience with Kramers' escape rate suggests that these integrals
should be evaluated (numerically) exactly \cite{Edholm79a},
particularly in situations such as that of Fig.
\ref{fig:N0_vs_T}(a) when the peaks of $\peq{N_{0}}$ are
well-defined but not extremely sharp. The rate coefficient
\begin{equation}
\Gamma=\w{N_{0}^{\text{max}}}{N_{0}^{\text{max}}-1}{\gets}\approx
\w{N_{0}^{\text{max}}}{N_{0}^{\text{max}}-1}{\to}
\end{equation}
is the only unknown quantity, thermalization-scheme-dependent and
beyond the scope of this paper to compute. Assuming the
temperature dependence of $\Gamma$ is slight next to that of
$\chi$, studying Eq. (\ref{chi}) shows that $\chi(\tau)$ has a
peak, located near $\tau_{\text{c}}$ and approaching it as
$\varepsilon$ decreases. For parameters as in Fig.\
\ref{fig:N0_vs_T}(a), $\chi(\tau_{\text{c}}) \approx
70,000\,\Gamma^{-1}$. For these parameters $\chi(\tau)$ does not
have a very sharp peak, but it acquires one with decreasing
$\varepsilon$.
\paragraph*{Conclusions.} We have presented an exactly solvable
semi-phenomenological
theory for the thermodynamics of harmonically trapped attractive
1D bosons. Our principal result is that this system has a
discontinuous no-soliton--soliton phase transition. We computed
the critical temperature and produced a phase diagram universal
with respect to experimental details such as trapping frequencies,
atomic mass, and the 3D scattering length. We also described the
major scaling laws governing the condensation time, showing it to
be exponentially long as one approaches the thermodynamic limit.
As a function of temperature, the transition time near the
critical temperature is given by a peaked curve, which we
determined up to a thermalization-scheme-dependent prefactor.

  Experiments thus far \cite{Salomon,Hulet} make
solitons by first condensing repulsive atoms and then switching
the sign of the coupling constant using the Feshbach resonance.
This produces certain excited solitonic states that are not states
of thermal equilibrium. In contrast, to observe the phase
transition considered in this paper one would need to produce
clouds of attractive 1D atoms in states of thermal equilibrium at
various temperatures. While challenging, experiments like this are
probably already feasible. Conceptually the simplest strategy may
be to perform experiments as in \cite{Salomon,Hulet}, but wait
long enough for the system to equilibrate before taking
measurements.
\begin{acknowledgments}
The authors are grateful to C. Salomon and R. G. Hulet for
enlightening discussions on the subject. Work of V.\ D.\ and M.\
O.\ was supported by the NSF grants \textit{PHY-0070333} and
\textit{PHY-0301052}. M.\ O.\ appreciates financial support by NSF
through the Institute for Theoretical Atomic and Molecular
Physics, Harvard Smithsonian Center for Astrophysics. C.\ H.\
thanks Laboratoire Kastler-Brossel for their hospitality during
his one-year visit. Laboratoire Kastler-Brossel is an
\textit{unit\'{e} de recherche de l'Ecole Normale Sup\'{e}rieure
et de l'Universit\'{e} Pierre et Marie Curie, associ\'{e}e au
CNRS}.
\end{acknowledgments}
%

\end{document}